\documentclass[aps,prl,reprint,superscriptaddress,longbibliography]{revtex4-2}
\usepackage[T1]{fontenc}
\usepackage[utf8]{inputenc}
\usepackage{graphicx}
\usepackage{amsmath,amssymb}
\usepackage[english]{babel}
\usepackage[dvipsnames]{xcolor}
\usepackage{pifont}
\usepackage{physics}
\usepackage{hyperref}
\newcommand{\prlsec}[1]{\noindent\textit{#1 ---}\hspace{0.5em}}
\usepackage[percent]{overpic}
\usepackage{xcolor} % in your preamble

\usepackage{etoolbox}

\makeatletter
\pretocmd{\bibsection}{\vspace*{-12pt}}{}{}  % space BEFORE the rule
\apptocmd{\bibsection}{\vspace*{-12pt}}{}{} % space AFTER the rule
\makeatother

\begin{document}

\title{Mechanism Behind the Recombination Requirement for\\Benign Termination of Relativistic Electron Beams}

\author{G.~Su}
\email{gs3416@columbia.edu}
\affiliation{Columbia University in the City of New York, USA}
%\affiliation{Columbia Fusion Research Center, Columbia University, New York, USA.}

\author{C.~F.~B.~Zimmermann}
\affiliation{Columbia University in the City of New York, USA}
%\affiliation{Columbia Fusion Research Center, Columbia University, New York, USA.}

\author{C.~Paz-Soldan}
\affiliation{Columbia University in the City of New York, USA}
%\affiliation{Columbia Fusion Research Center, Columbia University, New York, USA.}

\author{M.~H\"olzl}
\affiliation{Max-Planck-Institut f\"ur Plasmaphysik, Garching, Germany}

\author{P. Aleynikov}
\affiliation{Max-Planck-Institut f\"ur Plasmaphysik, Greifswald, Germany}

\begin{abstract}
\noindent We present a first-principles explanation of the recombination requirement for benign termination of relativistic electron (RE) beams in tokamaks. Kinetic modeling including neutrals shows that the injection of neutrals over a finite quantity window, together with recombination, increases bulk resistivity. Nonlinear MHD simulations using the JOREK code demonstrate that this preferentially amplifies edge tearing modes, producing a more stochastic edge magnetic field during RE deconfinement, resulting in a larger RE wetted area. We identify resistivity, not the free electron density, to govern access to benign termination. This provides the first broadly applicable and experimentally consistent picture of the MHD mechanisms behind the benign scenario, critical to its extrapolation to next-step devices.

\end{abstract}

\maketitle

\noindent The generation of relativistic electrons (REs) during tokamak disruptions remains a major challenge for the safe operation of future fusion devices~\cite{Loarte_2011,Lehnen_2013,ratynskaia2025runaway}. The most broadly demonstrated mitigation strategy is \textit{benign termination}, in which hydrogenic injection of specific quantities facilitates magnetohydrodynamic (MHD) instabilities that redistribute the RE beam heat load over a large wetted wall area, avoiding localized damage. Since its first demonstration on DIII-D~\cite{pazsoldan2019ppcf}, this scenario has been validated on multiple devices, including JET, ASDEX Upgrade (AUG), and TCV~\cite{reux2021prl,pazsoldan2021nf,sheikh2024ppcf}. 
A key uncertainty is the role of plasma density, which must decrease to recombination levels after injection~\cite{Hollmann2023,sheikh2024ppcf,Hoppe2025}. This has been attributed to density scaling of ideal MHD timescales. However, prior work shows the relevant instabilities involve resistive timescales, and benign and non-benign terminations cannot be distinguished by growth rates alone~\cite{Zimmermann_NF_2026}. Thus far, the link between material injection limits, recombination, and MHD dynamics remained unresolved, presenting a major obstacle for extrapolation to next-step devices. In addition, a general framework for the underlying MHD dynamics is lacking. Majority of previous modeling focused on double tearing modes \cite{bandaru2021magnetohydrodynamic,Nardon_2023}, which are not representative of most terminations in JET or of any in DIII-D \cite{Zimmermann_NF_2026}. Ref. \cite{Bandaru_2024} investigates the effect a 2/1 tearing mode (TM) but does not account for the involvement of kink modes which recent results identify as likely drivers for benign termination in current machines~\cite{Zimmermann_NF_2026}. Moreover, these simulations results show MHD evolution on timescales much slower than typical experimental observations. 

In this device-agnostic work, we combine kinetic collision modeling including neutrals with nonlinear extended-MHD simulations using the JOREK code~\cite{Hoelzl_2021, Hoelzl_2026} to study how resistivity and density affect the nonlinearly coupled evolution of internal kinks (IK) and TMs. We observe that increased resistivity preferentially amplifies edge tearing modes, leading to a transition from inside-out (non-benign) to outside-in (benign) stochastization structure. This results in drastic differences in the stochasticity of the magnetic field at the time of RE deconfinement, which we demonstrate to govern the RE wetted area. The MHD modeling is shown to be consistent with experiment in terms of mode structure, timescales, and resistivity. These results provide, for the first time, a self-consistent explanation of the recombination requirement for benign termination observed in present machines, offering critical guidance for extrapolation to future devices.

%%%%%%%%%%%%%%%%%%%%%%%%%%%%%%%%%%%%%%%%%%%%%%%%%%%%%%%%%%%%%%%%%%%%%%%%%%%%%%%%%%%%%

\prlsec{Kinetic modeling} A typical benign termination relies on hydrogenic injection into an already formed RE beam~\cite{reux2021prl, pazsoldan2021nf,sheikh2024ppcf}. Within a limited range of injected hydrogen quantities, this triggers recombination of the cold bulk plasma \cite{Hollmann2023,Hoppe2025}, leading to a substantial reduction in the free-electron density \(n_e\). In such conditions, where the neutral density can be orders of magnitude higher than \(n_e\), a fully ionized-plasma Spitzer resistivity model is not appropriate, since it is necessary to account for momentum transfer due to electron-neutral collisions~\cite{Vallhagen_2020}.
In a simplified description, the resistivity can be expressed in terms of an effective electron momentum transfer frequency as $\eta =  m_e \nu / (n_e e^2)$, where $\nu$ is understood as a kinetic-theory-based collision frequency. In a fully ionized plasma, the dominant contribution arises from Coulomb collisions with ions, $\nu_{ei} = n_e \sigma_{ei} v_{T_e} \sim n_e T_e^{-3/2}$, where $\sigma_{ei}$ is the Coulomb collision cross-section and $v_{T_e}$ the electron thermal speed. The explicit density dependence cancels, leading to the Spitzer $\eta$, which depends primarily on $T_e$. In a partially ionized plasma, an additional contribution from electron-neutral elastic collisions must be included, $\nu_{en} = n_0 \sigma_{en} v_{T_e}$ where $n_0$ is the neutral density and $\sigma_{en}$ is the electron-neutral momentum-transfer cross section, as tabulated in the literature~\cite{Golden_1966} and found to be identical for H and D. The total collision frequency can then be approximated as $\nu_{\text{tot}} = \nu_{ei} + \nu_{en} = v_{T_e} \left( n_e \sigma_{ei} + n_0 \sigma_{en} \right)$,
which gives an $\eta$ that includes both Coulomb and neutral contributions, 
\begin{equation*}
    \eta \sim v_{T_e} \left( \sigma_{ei} + \frac{n_0}{n_e} \sigma_{en} \right),
\end{equation*}
up to constant factors. This shows that the neutral term becomes large in regimes with low ionization fraction, resulting in a significant enhancement of $\eta$ near the recombination point.
\begin{figure}[]
    \centering
    \includegraphics[
        width=0.95\linewidth,
        trim={8 8 5 5},
        clip
    ]{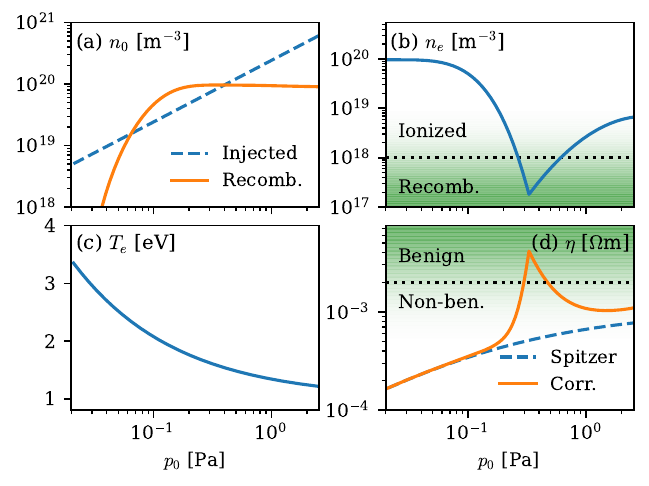}
    \caption{Illustrative kinetic modeling of neutral-induced recombination and its impact on bulk plasma resistivity $\eta$ as a function of the effective neutral pressure $p_0$. (a) Volume-averaged neutral density inferred from $p_0$ (blue, dashed) and increase in neutral density due to recombination (orange, solid). (b) Free electron density $n_e$ (blue) showing a recombination-driven drop of nearly two orders of magnitude, adapted from DIII-D \cite{Hollmann2023}. (c) Electron temperature $T_e$ adapted from TCV measurements \cite{Hoppe2025}. (d) Bulk plasma $\eta$: Spitzer-like model (blue, dashed) compared to a neutral-corrected model (orange, solid). The inclusion of neutrals leads to a peak around the access window for benign termination, which is marked by the green shaded regions.}
    \label{fig:kinetic_scan}
\end{figure}
To illustrate the impact of a neutral-corrected $\eta$ in this scenario, we consider a setup based on DIII-D-like neutral injection levels. The injected neutral inventory is related to an effective neutral pressure $p_0$ via the vessel volume, which serves as the control parameter in Fig.~\ref{fig:kinetic_scan}. This pressure defines a volume-averaged neutral density, see the blue curve in Fig. (a). Increasing $p_0$ drives strong recombination of the free electron density $n_e$ (Fig. (b)) due to neutral heat transport \cite{Hollmann2023}. The shown curve closely reproduces the measured data from DIII-D in Ref.~\cite{Hollmann2023} and is consistent with comparable TCV data \cite{Hoppe2025}. The recombination reduces $n_e$ by nearly two orders of magnitude and leads to a concurrent increase in neutral density, see the orange curve in Fig. (a). Electron temperature values $T_e$ are adopted from TCV measurements~\cite{Hoppe2025} (Fig. (c)). The resulting $\eta$ is shown in Fig. (d): a Spitzer-like model (blue) yields a smooth variation governed solely by $T_e$. In contrast, including neutral effects, based on adding the $n_0$ in Fig. (a) and setting them in relation to the $n_e$ from Fig. (b), produces a pronounced peak, with $\eta$ exceeding the Spitzer value by approximately an order of magnitude. Notably, this peaking occurs in the same parameter window where $n_e$ collapses and benign termination is observed~\cite{pazsoldan2021nf,Hollmann2023,sheikh2024ppcf,Hoppe2025}. This leads to the hypothesis that $\eta$ governs the underlying physics of the benign termination outcome.

This modeling is also consistent with the existence of an upper limit in injected quantity, beyond which $\eta$ decreases due to reionization of the background plasma \cite{Hollmann2023,Hoppe2025}. However, the resistivity beyond the upper limit does not return to the lowest values observed at low neutral pressure due to the intermediate density. This may indicate that, in certain scenarios, benign termination can also be achieved at higher neutral pressures, as observed in individual discharges on DIII-D~\cite{pazsoldan2021nf}. The influence of temperature assumptions in this calculation is minor: variations in $T_e$ shift the Spitzer curve uniformly but cannot account for the observed peaking or transition behavior. In particular, a Spitzer-like $\eta$ would require temperatures far below the ionization threshold to reach the $\eta$ values expected to be relevant for a benign scenario. It should be noted that radiation sources and interactions between the REs and the background plasma, as reioniziation discussed e.g. in Refs. \cite{Hollmann2023,Hoppe2025}, are not explicitly modeled. However, these effects are implicitly captured through the use of experimentally measured density and temperature values.

%%%%%%%%%%%%%%%%%%%%%%%%%%%%%%%%%%%%%%%%%%%%%%%%%%%%%%%%%%%%%%%%%%%%%%%%%%%%%%%%%%%%

\prlsec{MHD modeling} Kinetic modeling predicts a strong peak in neutral-corrected bulk resistivity, \(\eta\), within the recombination window, reaching $\sim 10^{-3}\,\Omega\mathrm{m}$, consistent with the transition between benign and non-benign termination. As shown previously~\cite{Helander_2007,Liu_2021,liu2026hybrid}, TM stability in the presence of REs is set by the \(\eta\), since REs are effectively collisionless.
Nonlinear extended-MHD simulations are performed using the \textsc{JOREK} code~\cite{Hoelzl_2021, Hoelzl_2026} to investigate the role of \(\eta\) in driving magnetic field stochastization conducive to benign termination. Background temperature and density are assumed to be constant due to the low thermal pressure of the cold, recombined plasma. The plasma current is assumed to be entirely carried by REs, treated as a separate fluid with an advection speed of 0.1\(c\) and coupled self-consistently through the current evolution~\cite{bandaru2021magnetohydrodynamic}. The fluid RE model is not used to reproduce RE transport with high fidelity, but rather to accurately capture MHD dynamics at high \(\eta\), as the collisionless RE population prevents resistive current decay expected in standard MHD. To capture edge stochastization without artificial influence of ideal wall boundary conditions, JOREK is coupled to the STARWALL resistive wall model~\cite{Holzl2012_STARWALL}. Resistivity is prescribed as a spatially uniform, time-independent parameter for each case.

The initial plasma configuration is based on typical parameters of benign termination experiments in medium size tokamaks like DIII-D or AUG: toroidal field \mbox{\(B_T = 2.2\,\mathrm{T}\)}, RE current \(I_{RE} = 0.72\,\mathrm{MA}\), major radius \(R_0 = 1.37\,\mathrm{m}\), minor radius \(a=0.4\,\mathrm{m}\) and nearly circular cross-section. We initialize the simulations after the RE beam has formed but just prior to the terminating instability. Experimental and linear MHD studies~\cite{Zimmermann_NF_2026} indicate the instability occurs after the system becomes unstable to internal kinks, motivating the choice of an edge safety factor of \(q_{\text{edge}} = 2.1\) and an internal inductance of \(l_i = 1.18\), corresponding to an IK unstable safety factor on axis of \(q_0 = 0.85\). Two reference cases are considered: low (\(\eta = 10^{-4}\,\mathrm{\Omega m}\), non-benign) and high (\(\eta = 10^{-2}\,\mathrm{\Omega m}\), benign) resistivity at fixed density \(n_e = 10^{18}\,\mathrm{m^{-3}}\).

A sensitivity study scanning toroidal harmonics up to \(n=5\) shows that the same \(n=1\) and \(n=2\) modes dominate the dynamics across all values of \(\eta\). Growth rate analysis shows that the 1/1 IK drives a 2/1 TM through toroidal coupling, with both modes exhibiting identical growth rates. Nonlinear three-wave coupling~\cite{CoelhoNLCouplingPoP} between these modes subsequently generates a 3/2 TM, which grows at twice the rate of the coupled 1/1 IK and 2/1 TM. This mode structure is consistent with synchrotron measurements on DIII-D~\cite{marini2024runaway}.

\begin{figure}
    \centering
    \includegraphics[width=0.95\linewidth]{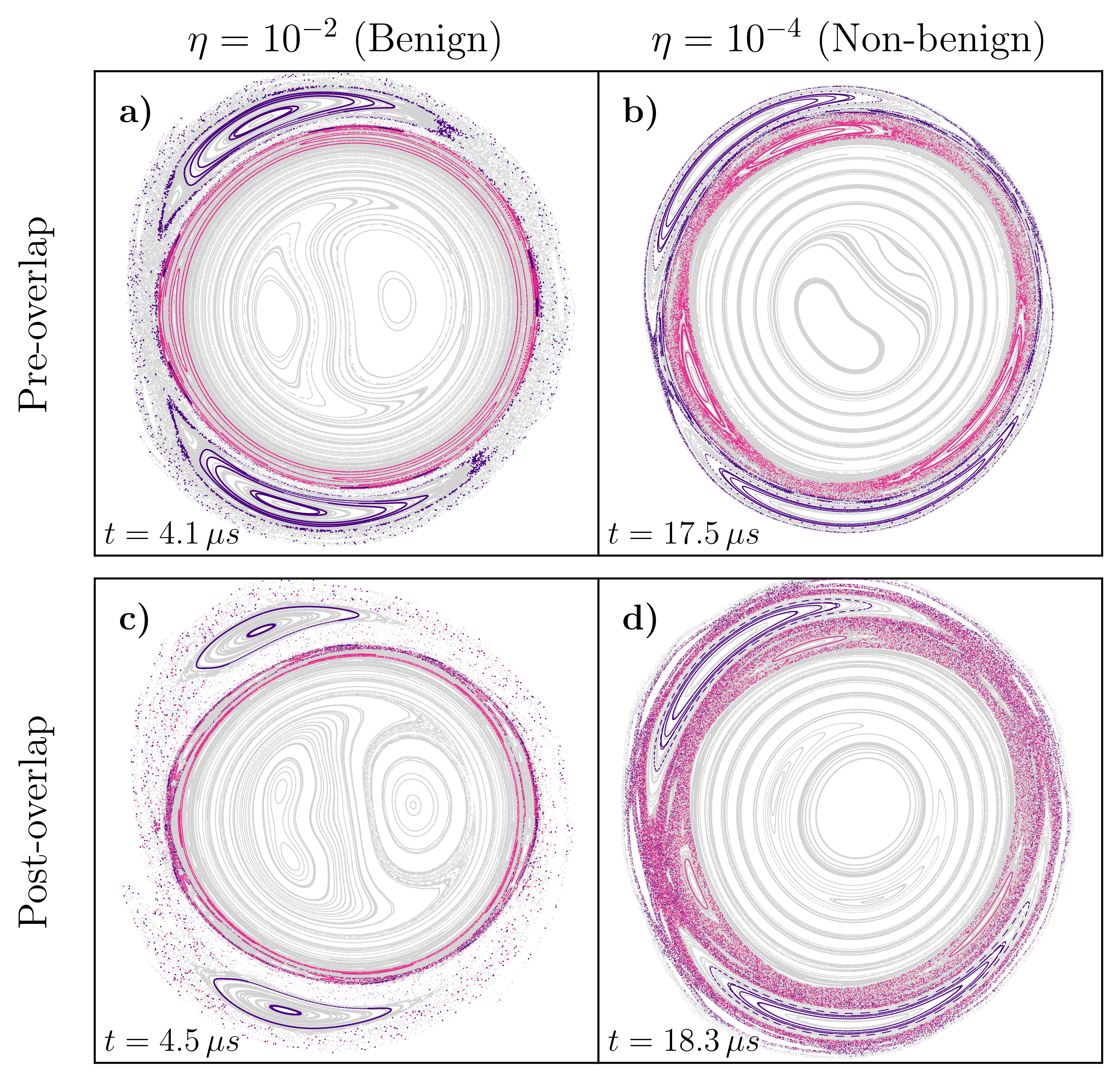}
    \caption{Poincar\'e plots illustrating magnetic field-line stochastization. Exterior (purple) and interior (pink) field-lines are influenced by the \(2/1\) and \(3/2\) TM, respectively. Rows correspond to temporal snapshots before mode-overlap (upper) and shortly after mode-overlap (lower). Columns show resistivity cases: (left) \(\eta = 10^{-2}\,\Omega\mathrm{m}\), (right) \(\eta = 10^{-4}\,\Omega\mathrm{m}\).} 
    \label{fig:poincare}
\end{figure}

The magnetic islands formed by these TMs as well as the perturbation of the core due to the 1/1 IK are clearly visible in Fig.~\ref{fig:poincare}(a,b) (purple: 2/1 TM, pink: 3/2 TM, gray 1/1 kink), which shows the magnetic field topology just before the islands from the two TMs grow large enough to overlap. We see that prior to TM-overlap, radially localized field stochastization develops gradually. 
Crucially, the nonlinear growth rates of the $2/1$ and $3/2$ TMs show distinct scaling with $\eta$. In the $\eta = 10^{-4}\,\mathrm{\Omega\,m}$ (non-benign) case, the $3/2$ TM grows faster than the $2/1$ TM, producing a strongly stochastic core and a weakly stochastic edge at mode-overlap. Following overlap, see Fig.~\ref{fig:poincare}(d), an extended stochastic region forms, allowing interior REs to immediately access field lines that intersect the wall. This sets a finite time window for further stochastization, as REs will rapidly escape the plasma once a path is made available. In this case, the edge stochastization remains weak throughout the simulated time window, resulting in a narrow deposition on the wall as shown later.
In contrast, for the $\eta = 10^{-2}\,\mathrm{\Omega\,m}$ (benign) case, shown in Fig.~\ref{fig:poincare}(c), the $2/1$ TM dominates over the $3/2$ TM. This inverts the structure so that the edge becomes more stochastic than the core. REs from the $3/2$ region are now released into an already well stochastic edge region, providing a broad set of escape paths and increasing the wetted area. The weaker stochasticity at low $\eta$ is reflected by the much higher point density, as field-lines undergo many more toroidal transits before loss.
Shortly after the TM overlap, the $3/2$ TM overlaps with the $1/1$ island associated with the IK, producing global stochastization and enabling deconfinement of the remaining REs. While the detailed overlap dynamics depend on the specific $q$-profile and the influence of kinetic RE effects on the MHD evolution, we identify the level of edge stochasticity at the onset of global stochasticity as the primary factor governing benign termination. This is determined by the relative growth rates of the modes, which, due to their coupled evolution, are largely insensitive to the $q$-profile and kinetic RE effects \cite{Helander_2007, liu2026hybrid}.

\begin{figure}
    \centering
    \includegraphics[width=0.95\linewidth]{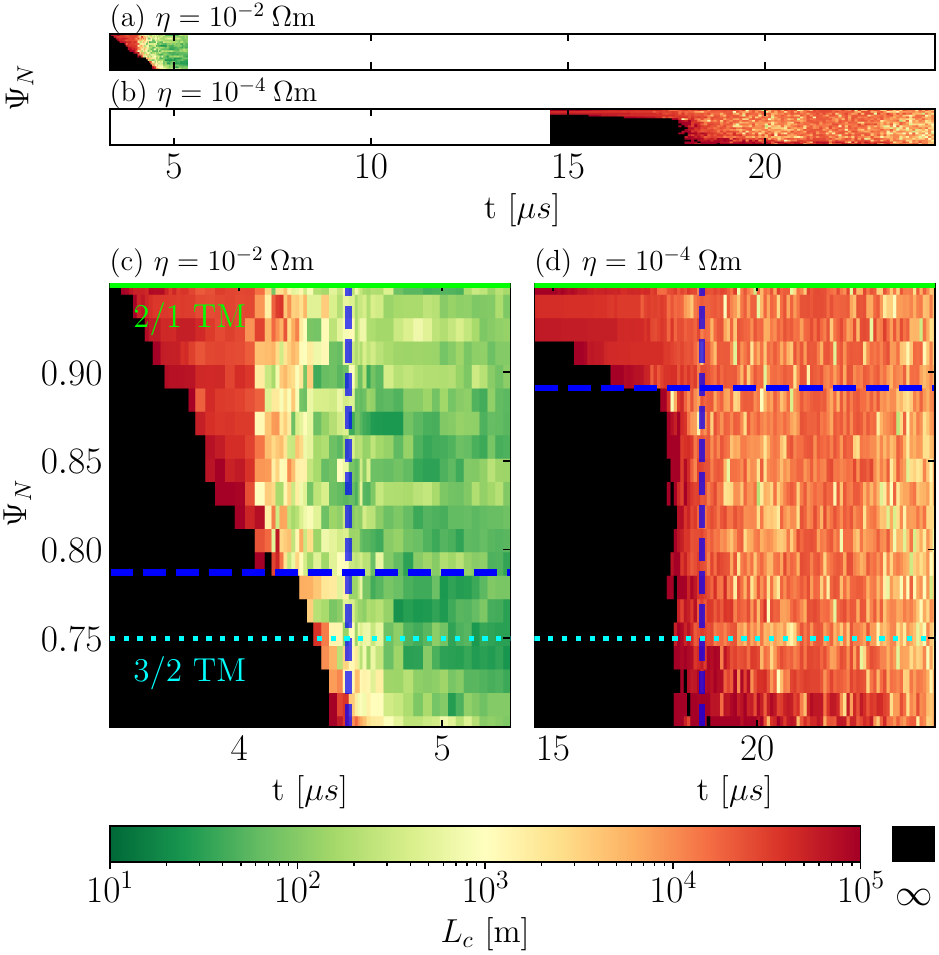}
    \caption{Spatiotemporal evolution of the poloidally averaged connection length $L_c$ (color map) for field-lines originating in the radial region between $\Psi_N=0.7$ and $0.95$ (y-axis). Black indicates \(L_c = \infty\) (confined). Simulations are shown for resistivities $\eta=10^{-2} \,\Omega\mathrm{m}$ (a,c) and $10^{-4}\,\Omega\mathrm{m}$ (b,d). The same data is plotted twice: At the top, a common time axis is used (a,b) highlighting a factor of six difference in time scale of the event. At the bottom, a case-adapted time axis is used spanning the relevant MHD event for each case (c,d). The horizontal lines denote position of the 2/1 rational surface (green, solid), 3/2 rational surface (cyan, dotted) and the radial position at which TM-overlap arises (dark blue, dashed). The vertical line denotes the onset of global stochasticity.}
    \label{fig:connection_length}
\end{figure}

For a more quantitative measure of local stochasticity, we use the connection length $L_c$, defined as the distance a field line travels from its starting point to the wall. $L_c$ is infinite for confined field lines and decreases as stochasticity increases. Figure~\ref{fig:connection_length} shows the evolution of $L_c$ (colormap) as a function of radial position (measured by \(\Psi_N\)) (y-axis) and time across the mode-overlap phase (x-axis). We focus on the edge region ($\Psi_N>0.7$), as it ultimately determines the RE wetted area. As we are primarily interested in the radial differences in stochasticity, $L_c$ is computed as the harmonic mean over ten field lines evenly spaced in poloidal angle at fixed radius.  The same two reference cases with \(\eta=10^{-2} \, \mathrm{\Omega m} \) (a,c) and \(\eta=10^{-4} \, \mathrm{\Omega m} \) (b,d) are presented.
For comparing time scales, Figs. (a,b) present both cases on a shared time axis. The MHD activity occurs earlier and evolves faster with increased \(\eta\), see Fig. (a). The general time scale of the instabilities of order 1-10 $\mathrm{\mu s}$ matches with the experimentally observed instability time scales, see Ref.~\cite{Zimmermann_NF_2026}. Figures (c) and (d) provide a zoomed in version of each case and provide a clearer comparison of resulting stochasticity. The picture is consistent with the observations from the Poincar\'e plots. In both cases, the $2/1$ TM stochasticizes the plasma edge, leading to finite \(L_c\) prior to overlap, while the \(3/2\) island remains initially confined despite local stochastization. The horizontal dashed, dark-blue line marks the radial location of TM-overlap, while the solid, green and dotted cyan lines marks 2/1 and 3/2 rational surfaces. We see that the TM-overlap position skews towards the $3/2$ in the \(\eta=10^{-2}\,\mathrm{\Omega m}\) case, since the \(2/1\) TM dominates over the \(3/2\), thereby initiating overlap. This is further supported by the sloped deconfinement contour (transition from black to color), indicating limited penetration of the \(3/2\) mode prior to deconfinement. Conversely, in the \(\eta=10^{-4}\,\mathrm{\Omega m}\) case, the more exterior overlap position indicates that the \(3/2\) mode initiates overlap, producing a sharp, nearly vertical drop in the deconfinement contour as a large area of pre-stochasticized region suddenly gains access to the plasma edge. 
The vertical dashed line indicates the onset of global stochasticity produced by the overlapping of $3/2 $ TM and the $1/1$ island. In the \(\eta=10^{-2}\,\mathrm{\Omega m}\) case, the dominant \(2/1\) TM has already sufficiently stochastized the field, as reflected by the low \(L_c\) values (green/yellow). As a result, REs released from the core traverse a strongly stochastic field on their way out, likely leading to a large wetted area, and thus benign termination. In contrast, for \(\eta=10^{-4}\,\mathrm{\Omega m}\), the \(3/2\) TM triggers the onset of global stochasticity before the \(2/1\) mode can stochastize the edge strongly. Consequently, \(L_c\) remains $2-3$ orders of magnitude higher (orange/red), likely leading to a more localized deposition pattern and a smaller wetted area.

\begin{figure}
    \centering
    \begin{overpic}[width=0.95\linewidth]{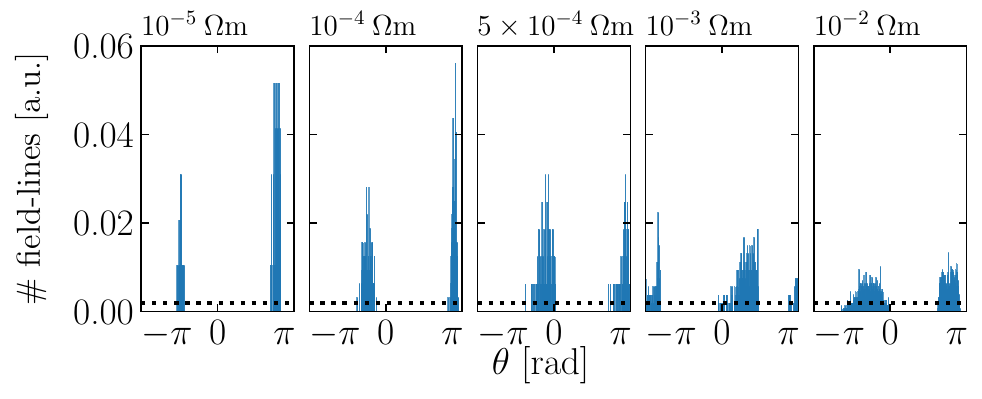}
        \put(0,39){\scalebox{0.9}{(a)}}
    \end{overpic}
    
    \vspace{1.0em}
    
    \begin{overpic}[width=0.95\linewidth]{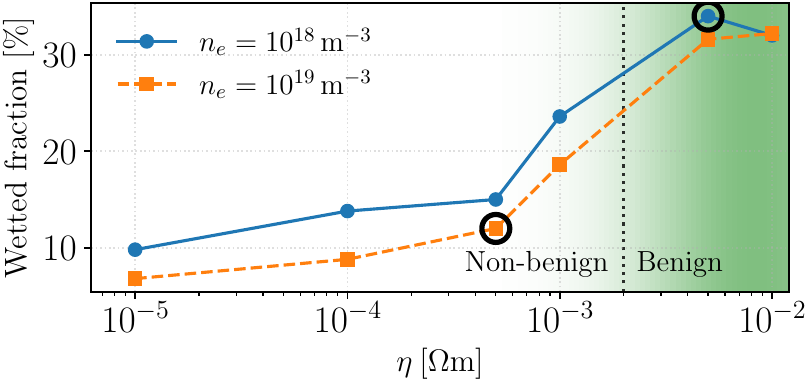}
        \put(0,50){\scalebox{0.9}{(b)}}
    \end{overpic}
    
    \caption{(a) Histogram with 500 equally spaced bins showing the distribution of the poloidal position interior field-lines intersect an artificial external surface at \(\Psi_N=1.1\) accumulated over the deconfinement window for different \(\eta\), simulated with \(n_e = 10^{18}\,\mathrm{m^{-3}}\). Field-line counts per bin have been normalized to sum to 1. (b) Fraction of poloidal area wetted, computed as the fraction of bins with more than 0.2\% (dotted line in (a)) of the total number of field-lines. Green shading shows the benign termination $\eta$ range from Fig.~\ref{fig:kinetic_scan}.}
    \label{fig:theta_dispersion}
\end{figure}

Having established the mechanism underlying magnetic stochasticity, we next assess its impact on the RE wetted area. We do so using the poloidal distribution of the termination positions of field-lines that escape after the onset of global stochasticity, which is computed via intersections with an artificial surface outside the plasma ($\Psi_N=1.1$). We focus on the poloidal direction, which has been shown to be the dominant dimension along which the wetted area varies~\cite{pazsoldan2021nf,Bergström_2025}. A fully quantitative heat-load calculation would require kinetic RE tracing and a detailed wall model, which is beyond the scope of this work. To reduce sensitivity to short-time fluctuations, we collect the distribution from $t_{GS}$ (onset of global stochasticity) to $t_{GS}+t_{RD}$, where $t_{RD}=0.7\,\mu s$ is the Rosenbluth diffusion time~\cite{RosenbluthPRL1978} calculated with $\delta B/B=0.02$, typical of benign termination experiments~\cite{Zimmermann_NF_2026,pazsoldan2021nf}. Owing to the dominance of parallel transport for REs, the calculated quantities serve as a proxy to illustrate the effect of the modified topology from the perspective of the REs.
Figure~\ref{fig:theta_dispersion}(a) shows a histogram of this angular distribution for five different values of \(\eta\), all simulated with \(n_e = 10^{18}\,\mathrm{m^{-3}}\). The x-axis is the terminating poloidal angle and the y-axis is the fraction of total field-lines terminating there. For the lowest $\eta$ value, two distinct narrow spikes are seen, indicating a very focused deposition. With increasing \( \eta \), these peaks are smoothed out, indicative of a larger wetted area and benign termination.
Figure~\ref{fig:theta_dispersion}(b) provides a quantitative measure of the fraction of the poloidal surface wetted by evaluating the fraction of the bins with more than 0.2\% (dotted line in Fig.~\ref{fig:theta_dispersion}(a)) of the total number of field-lines. We see a clear trend of increasing wetted area with increasing \(\eta\). We observe that the $\eta$ dependency increases strongly for \(\eta>5\times10^{-4}\), which is well within the \(\eta\) range for benign termination identified by the kinetic modeling (green shaded region). This transition is found to coincide with the \(\eta\) value where the rate of local stochastization from the $2/1$ TM exceeds that of the $3/2$ TM.  
The orange line shows the results from the same range of \(\eta\) values, but simulated with \(n_e = 10^{19}\,\mathrm{m^{-3}}\) instead, corresponding to the non-recombined limit. All observed trends and dynamics discussed closely follow the results for \(n_e = 10^{18}\,\mathrm{m^{-3}}\). The wetted area is found to be $\sim3$~\% lower on average. This difference is likely due to the longer ideal MHD timescales, which reduces the \(2/1\) TM growth rate. However, the variation in wetted fraction due to \(n_e\), especially for \(\eta\) values close to the experimental threshold is significantly smaller than that due to \(\eta\), and is far smaller than that seen in experiment. This result shows that it is nonlinear resistive effects which govern benign termination physics, and change in density itself only plays a secondary role.
This isolation of the effects of \(n_e\) and \(\eta\) is of course artificial as shown by the kinetic modeling. Restricting to physically consistent \(n_e\)–\(\eta\) pairs selected from Fig.~\ref{fig:kinetic_scan} (circled points), our model finds a 2.8 times increase in wetted area across the benign threshold, which is quantitatively consistent with experimental measurements~\cite{sheikh2024ppcf, pazsoldan2021nf}. 

%%%%%%%%%%%%%%%%%%%%%%%%%%%%%%%%%%%%%%%%%%%%%%%%%%%%%%%%%%%%%%%%%%%%%%%%%%%%%%%%%%%%%

\prlsec{Discussion and Conclusion} We have established a self-consistent framework linking neutral-induced recombination to benign termination through its effect on bulk resistivity and the resulting nonlinear resistive MHD dynamics. Kinetic modeling shows that recombination produces a non-monotonic increase in bulk resistivity, which governs IK–TM coupling. This results in stronger edge stochasticity at the time of RE deconfinement, ultimately broadening the RE wetted area and leading to benign termination. Preliminary analysis of RE deposition shows a 2.8 times increase in wetted area, which is consistent with the transition from benign to non-benign termination found in experiments \cite{sheikh2024ppcf, pazsoldan2021nf}. This emphasis on terminating edge stochasticity is consistent with the theoretical framework by Boozer and Punjabi \cite{Boozer_2016}.
This framework reproduces experimental observations in current machines for the first time, including time scales, mode structures, and the relevant density/resistivity range, and offers an explanation for the existence and variability of density limits observed across devices.
Our results indicate that benign termination is a smooth transition in stochasticity rather than a sharp threshold, consistent with TCV measurements~\cite{sheikh2024ppcf}. Although kinetic RE effects are not included, prior work in Refs.~\cite{Helander_2007,liu2026hybrid} shows limited impact on growth rates, whereas island sizes are expected to increase, exaggerating the dominance of the 3/2 and 2/1 TM at low and high \(\eta\) respectively, hence further increasing the observed variation in wetted area.
Besides \(\eta\), parameters such as shaping, $q_a$, $q_0$, and $l_i$ influence the MHD dynamics and can be used to optimize stochastization for benign termination, e.g., via optimizing the current profile. In particular, higher $l_i$ steepens current gradients and accelerates TM growth, potentially explaining the absence of benign termination on JET at $q=2$, where an IK is seeded but $l_i$ values are consistently lower than in \mbox{DIII-D~\cite{Zimmermann_NF_2026}}. Alternatively, the resistivity requirement may be relaxed by avoiding the IK, and instead reaching termination through external kinks, as also observed on JET, where a similar neutral injection window was not observed for pre-disruptive plasma currents $\lessapprox1.5$ MA. Firm conclusions for JET are still pending due to the challenge of extracting reliable density measurements.
While suppression of magnetic to kinetic energy conversion is important for mitigating RE loads \cite{Loarte_2011}, we do not address this process here, as it requires kinetic modeling \cite{Bergström_2025} and is left for future work. Instead, we note that the timescales for magnetic field stochastization observed here is much faster than the resistive timescale for magnetic to kinetic energy conversion, which has been shown to result in negligible conversion rates \cite{Martin-Solis_2014}.
Finally, this framework opens pathways for control and extrapolation, including the use of resonant magnetic perturbations to assist stochastization and applications to future devices such as ITER and SPARC. Initial scoping for SPARC, for example, suggests that despite the absence of strong recombination \cite{Hollmann2023} and, consequently, a smaller drop in background density, the neutral content in the vessel alone could raise the resistivity to levels sufficient to trigger benign termination, owing to a sufficiently large $n_0/n_e$. More broadly, the presented results enable new strategies to optimize and reliably achieve benign termination in future, high-current fusion devices.

\prlsec{Acknowledgments} The authors would like to thank \mbox{E. Hollmann}, F. Antlitz, A. Cathey, H. Bergstr\"{o}m and \mbox{R. Sparago} for valuable discussions. The first author would like to acknowledge co-funding from the Max Planck Institute for Plasma Physics. 
This work was supported by the U.S. Department of Energy, Office of Science, Office of Fusion Energy Sciences under Award(s) DE-SC0022270, and the U.S. Department of Energy FIRE Collaborative “Mitigating Risks from Abrupt Confinement Loss (MiRACL)” under contract numbers DE-AC02-09CH11466 and DE-AC05-00OR22725. The United States Government retains a non-exclusive, paid-up, irrevocable, world-wide license to publish or reproduce the published form of this manuscript, or allow others to do so, for United States Government purposes.
This work has also been carried out within the framework of the EUROfusion Consortium, partly funded by the European Union via the Euratom Research and Training Programme (Grant Agreement No 101052200 EUROfusion). Views and opinions expressed are however those of the author(s) only and do not necessarily reflect those of the European Union or the European Commission. Neither the European Union nor the European Commission can be held responsible for them. 
%
% This material is based upon work supported by
% the U.S. Department of Energy, Office of Science, Office of Fusion Energy Sciences, under Award(s) DE-SC0022270. This report was prepared as an account of work sponsored by an
% agency of the United States Government. Neither the United States Government nor any
% agency thereof, nor any of their employees, makes any warranty, express or implied, or
% assumes any legal liability or responsibility for the accuracy, completeness, or usefulness
% of any information, apparatus, product, or process disclosed, or represents that its use
% would not infringe privately owned rights. Reference herein to any specific commercial
% product, process, or service by trade name, trademark, manufacturer, or otherwise does
% not necessarily constitute or imply its endorsement, recommendation, or favoring by the
% United States Government or any agency thereof. The views and opinions of authors expressed herein do not necessarily state or reflect those of the United States Government
% or any agency thereof.

\bibliographystyle{apsrev4-2}
\bibliography{main}
\end{document}